\documentclass[conference, final]{IEEEtran}

\newcounter{MYtempeqncnt}
\DeclareUnicodeCharacter{0306}{{\u \i}} 

\usepackage{amsmath,amssymb,amsfonts}
\usepackage{algorithmic}
\usepackage{graphicx}
\usepackage{textcomp}
\usepackage{xcolor}
\usepackage{dblfloatfix}
\usepackage{url}
\usepackage{hyperref}
\usepackage[font=small]{caption}
\usepackage{subfig} 
\hypersetup{
    colorlinks=true,
    linkcolor=blue,
    urlcolor=blue,
    filecolor=blue,
    citecolor=blue
    }

\usepackage{ifthen}
\makeatletter
\newcounter{IEEE@bibentries}
\renewcommand\IEEEtriggeratref[1]{%
  \renewbibmacro{finentry}{%
    \stepcounter{IEEE@bibentries}%
    \ifthenelse{\equal{\value{IEEE@bibentries}}{#1}}
    {\finentry\@IEEEtriggercmd}
    {\finentry}%
  }%
}
\makeatother

\usepackage{tikz}
\newcommand\copyrighttext{%
    \footnotesize \textcopyright 2025 IEEE. Personal use of this material is permitted. Permission from IEEE must be obtained for all other uses, in any current or future media, including reprinting/republishing this material for advertising or promotional purposes, creating new collective works, for resale or redistribution to servers or lists, or reuse of any copyrighted component of this work in other works.
    DOI: \href{https://doi.org/10.1109/PDP66500.2025.00060}{10.1109/PDP66500.2025.00060}}
\newcommand\copyrightnotice{%
    \begin{tikzpicture}[remember picture,overlay]
        \node[anchor=south,yshift=20pt] at (current page.south) {\fbox{\parbox{\dimexpr\textwidth-\fboxsep-\fboxrule\relax}{\copyrighttext}}};
    \end{tikzpicture}%
}

\usepackage[style=ieee,maxnames=8,minnames=8]{biblatex}
\usepackage{listings}
\def\BibTeX{{\rm B\kern-.05em{\sc i\kern-.025em b}\kern-.08em
    T\kern-.1667em\lower.7ex\hbox{E}\kern-.125emX}}

\begin{document}

\title{\texttt{BrahMap}: A scalable and modular map-making framework for the CMB experiments
}


\author{\IEEEauthorblockN{Avinash Anand\IEEEauthorrefmark{1}, Giuseppe Puglisi\IEEEauthorrefmark{2}} \\

\IEEEauthorblockA{\IEEEauthorrefmark{1}Dipartimento di Fisica, Università di Roma Tor Vergata, Via della Ricerca Scientifica, 1, 00133, Roma, Italy}
\IEEEauthorblockA{\IEEEauthorrefmark{1}INFN Sezione di Roma2, Università di Roma Tor Vergata, Via della Ricerca Scientifica, 1, 00133 Roma, Italy} 
\IEEEauthorblockA{\IEEEauthorrefmark{2}Dipartimento di Fisica e Astronomia, Università degli Studi di Catania, Via S. Sofia, 64, 95123, Catania, Italy}
\IEEEauthorblockA{\IEEEauthorrefmark{2}	INAF, Osservatorio Astrofisico di Catania, Via S.Sofia 78, I-95123 Catania, Italy}
\IEEEauthorblockA{\IEEEauthorrefmark{2}INFN, Sezione di Catania, Via S.Sofia 64, I-95123, Catania, Italy}
Email: \IEEEauthorrefmark{1}avinash.anand@students.uniroma2.eu, \IEEEauthorrefmark{2}giuseppe.puglisi@dfa.unict.it
}



\newcommand{\red}[1]{\textcolor{red}{#1}}

\newcommand{\green}[1]{\textcolor{green}{#1}}






\newcommand{\blue}[1]{\textcolor{blue}{#1}}

\newcommand{\lime}[1]{\textcolor{lime}{#1}}

\newcommand{\orange}[1]{\textcolor{orange}{#1}}


\maketitle

\copyrightnotice

\begin{abstract}
The cosmic microwave background (CMB) experiments have reached an era of unprecedented precision and complexity. Aiming to detect the primordial B-mode polarization signal, these experiments will soon be equipped with 10\textsuperscript{4} to 10\textsuperscript{5} detectors. Consequently, future CMB missions will face the substantial challenge of efficiently processing vast amounts of raw data to produce the initial scientific outputs - the sky maps - within a reasonable time frame and with available computational resources. To address this, we introduce \texttt{BrahMap}, a new map-making framework that will be scalable across both CPU and GPU platforms. Implemented in C++ with a user-friendly Python interface for handling sparse linear systems, \texttt{BrahMap} employs advanced numerical analysis and high-performance computing techniques to maximize the use of super-computing infrastructure. This work features an overview of the \texttt{BrahMap}'s capabilities and preliminary performance scaling results, with application to a generic CMB polarization experiment.
\end{abstract}


\section{Introduction to CMB map-making}

Cosmic inflation provides an elegant solution to a number of open problems in cosmology, and its several predictions have been tested as well. One distinct prediction of inflation is the production of stochastic gravitational waves (GW) that has not been observed yet. The B-mode polarization of cosmic microwave background (CMB) from recombination is predicted to carry the imprints of these stochastic GW from the early Universe \cite{cmb1,cmb2,cmb3}. A number of space and ground-based experiments are therefore planned for the upcoming decade to probe the B-mode polarization of CMB \cite{experiment1, experiment2, experiment3}. These experiments will employ thousands of detectors sampling the signal at very high frequencies, with data acquisition up to several hundred terabytes. Before proceeding with scientific analysis, these data must be reduced to sky maps in order to produce the intended scientific outcome of the experiment.

Map-making is a data reduction problem that aims to reduce the time-ordered data (TOD) to sky-signal maps. The sky maps obtained after the map-making are processed through the component separation pipeline to obtain the maps of CMB and different galactic and extragalactic foregrounds. The process of map-making converts the time-domain data to the sky-pixel domain. It is an important stage of the data analysis pipeline where the time-domain artifacts can enter the sky-pixel domain. Since the time-domain artifacts have complex dependency on the scanning strategy, foregrounds, instrument calibration and systematics, etc., it is often hard to model them in sky-pixel domain. Therefore, it is crucial to perform the map-making while taking care of the time-domain artifacts. From a computational point-of-view, map-making involves the reduction of terabytes of data in the time-domain to ${\sim}100$ GB of data in the sky-pixel domain. It makes the map-making both a compute and a data intensive problem.

Although map-making may appear to be a one-time data reduction challenge, it is, in fact, tightly linked with instrument development. During the development phase of a CMB experiment, we study the flow-down of scientific mission requirements to instrument specifications, and vice versa. Map-making, which serves as a link between raw experiment data and the first scientific results, provides important feedback for evaluating instrument parameters and refining the end-to-end data analysis pipeline. As such, it plays a crucial role in every stage of the CMB experiment, from development to completion.


\section{Data model and formalism}

In a typical CMB experiment, each detector scans the sky continuously following a certain scanning strategy. As such, they aim to observe a sky pixel multiple times in order to obtain a high signal-to-noise ratio in each pixel. The data collected by each detector constitute a timeline instead of images.

Assuming that a detector observes a single sky-pixel at a time, the signal captured by a detector at time $t$ is given by

\begin{equation}
d_t = s_{p,t} + n_t \label{eq:data_model}
\end{equation}

where $s_{p,t}$ is the signal contribution from the sky pixel $p$ at time $t$ and $n_t$ is the total noise at time $t$. The noise component includes both un-correlated (white) and correlated noise. For a typical CMB polarization experiment, the sky signal can be seen as the sum of the contributions from different Stokes parameters of the sky:

\begin{equation}
s_{p,t} = I_{p,t} + Q_{p,t}\cos2\theta_t + U_{p,t}\sin2\theta_t \label{eq:sky_signal}
\end{equation}

\begin{figure*}[!b]
\small
\hrulefill
\vspace*{4pt}
\setcounter{MYtempeqncnt}{\value{equation}}
\setcounter{equation}{7} 
\begin{equation}
d = \begin{bmatrix}
d_1\\
d_2\\
d_3\\
\vdots\\
d_{N_{t-2}}\\
d_{N_{t-1}}\\
d_{N_{t}}
\end{bmatrix}\, P = \begin{bmatrix}
 \dots \! & \dots \! & \dots \! & \dots \! & \dots \! & 1 \! & \cos2\theta_{t_1} \! & \sin2\theta_{t_1} \! & \dots \! \\
 \dots \! &  1 \! & \cos2\theta_{t_2} \! & \sin2\theta_{t_2} \! & \dots \! & \dots \! & \dots \! & \dots \! & \dots \! \\
 \dots \! & \dots \! & \dots \! & \dots \! & \dots \! & 1 \! & \cos2\theta_{t_3} \! & \sin2\theta_{t_3} \! & \dots \! \\
 \dots \! & \dots \! & \dots \! & \dots \! & \dots \! & \dots \! & \dots \! & \dots \! & \dots \! \\
 \dots \! & \dots \! & \dots \! & \dots \! & \dots \! & 1 \! & \cos2\theta_{t_{N_{t-2}}} \! & \sin2\theta_{t_{N_{t-2}}} \! & \dots \! \\
 \dots \! &  1 \! & \cos2\theta_{t_{N_{t-1}}} \! & \sin2\theta_{t_{N_{t-1}}} \! & \dots \! & \dots \! & \dots \! & \dots \! & \dots \! \\
 \dots \! &  1 \! & \cos2\theta_{t_{N_t}} \! & \sin2\theta_{t_{N_t}} \!  & \dots \! & \dots \! & \dots \! & \dots \! & \dots \!
\end{bmatrix}\, s = \begin{bmatrix}
I_1\\
Q_1\\
U_1\\
\vdots\\
I_{N_p}\\
Q_{N_p}\\
U_{N_p}
\end{bmatrix}\, n = \begin{bmatrix}
n_1\\
n_2\\
n_3\\
\vdots\\
n_{N_{t-2}}\\
n_{N_{t-1}}\\
n_{N_{t}}
\end{bmatrix} \label{eq:define_matrices}
\end{equation}
\setcounter{equation}{\value{MYtempeqncnt}}
\end{figure*}

In this equation, $I$, $Q$ and $U$ are the Stokes parameters for the intensity and linear polarization, respectively. The angle $\theta$ indicates the projection angle of linear polarization as it is captured by the detector. Usually, it is the sum of the orientation angle of the polarization detector and the rotation of the polarization angle by different optical elements. For the sake of brevity, we consider here $\theta$ to be inclusive of all such components.

With sky signal and total signal defined in equations \eqref{eq:sky_signal} and \eqref{eq:data_model}, the collection of these equations for the entire TOD can be seen as an inhomogeneous system of linear equations. In this system of linear equations, the known quantities are $d_t$ and $\theta_t$ and the unknown quantities are $I_{p,t}$, $Q_{p,t}$, $U_{p,t}$ and $n_t$. We can write this inhomogeneous system of linear equations in matrix form as follows:

\begin{equation}
d = P\cdot s + n \label{eq:matrix_form}
\end{equation}

where $d$, $s$, and $n$ are signal, sky, and noise vector, respectively. The matrix $P$ on the other hand is called the pointing matrix and it projects each time sample to the sky pixel. If the number of sky pixels observed is given by $N_p$ and the total number of samples collected is given by $N_t$, the size of the pointing matrix will be $N_t \times 3N_p$. With the assumption that a detector captures only one sky pixel at a time, each row of the pointing matrix will contain only three non-zero elements. The three non-zero elements will be 1, $\cos\theta_{n_i}$ and $\sin\theta_{n_i}$ respectively in the columns corresponding to the observed pixel and in the row corresponding to their time index. A general explicit form of $d$, $P$, $s$, and $n$ is given in equation \eqref{eq:define_matrices}.

The matrix equation \eqref{eq:matrix_form} is a single equation with two known quantities - $d$ and $P$ - and two unknown quantities - $s$ and $n$. Our aim is to solve the matrix equation for the sky vector $s$. Assuming the noise vector to be Gaussian and stationary, we can write the linear solution for the sky vector $s$ as \cite{ml1, ml2}:

\begin{equation}
\hat{s} = \left(P^T W P\right)^{-1} P^T W d \label{eq:maxl_solution}
\end{equation}

where $W$ is any positive-definite symmetric matrix compatible with $P$. For the noise covariance given by $N$, setting $W = N^{-1}$ in equation \eqref{eq:maxl_solution} will provide a minimum-variance estimate or the Generalized Least Square solution (GLS) \cite{gls1} for the sky vector:

\begin{equation}
\hat{s} = \left(P^T N^{-1} P\right)^{-1} P^T N^{-1} d \label{eq:GLS}
\end{equation}

In the GLS equation, the matrix $P^T N^{-1} P$ has the typical dimension $10^6\times10^6$ or higher whereas the dimension of the noise covariance matrix $N$ depends on the length of the TOD, $N_t$, which could be larger than $10^{12}$. The inversion of these enormous matrices makes GLS a computationally challenging problem to solve. In general, we make some assumptions on the structures of noise covariance to approximate $N^{-1}$ and solve equation \eqref{eq:GLS} as a linear equation using iterative solvers like conjugate gradient method with suitable pre-conditioner:

\begin{equation}
  A\cdot x = b \label{eq:lineq}
\end{equation}

where

\begin{gather*}
  A = P^T N^{-1} P \\
  x = \hat{s} \\
  b = P^T N^{-1} d
\end{gather*}

To solve equation \eqref{eq:lineq} with PCG, a block-diagonal preconditioner based on the Jacobi preconditioner is commonly used:

\begin{equation}
M = \left(P^T diag(N)^{-1} P\right)^{-1} \label{eq:precond}
\end{equation}

Due to the structure of the pointing matrix $P$, the matrix $(P^T diag(N)^{-1} P)$ becomes a block-diagonal matrix with block size $3\times3$. Inverting this blocks-diagonal matrix is equivalent to inverting each $3\times3$ blocks individually, which can be easily computed numerically. So, if the diagonal elements of the noise covariance $N$ are known in advance, computing $M$ becomes a trivial exercise.

Implementations of GLS written in the past have been instrumental in map-making for various ballon-based CMB experiments and space missions like WMAP \cite{wmap} and \textit{Planck} \cite{planck}. Such implementations were successfully able to leverage the computational facilities to handle the data collected by the previous generation experiments. Though these implementations are still relevant, based on the projected data acquisition size, we estimate that they will require somewhere around ${\sim} 200,000$ (for a typical space-based mission) to ${\sim} 50,000,000$ (for a typical ground-based experiment) CPU hours to perform the map-making on data coming from future CMB experiments. If we include the map-making exercise that is performed during the development phase of these experiments, the computational cost for mission development will only increase. In light of future CMB experiments and the exponential growth in computational capabilities in the past decade, it is important that we revisit the map-making implementations and come up with the solution that can serve the computational demand of future experiments while taking full advantage of the new distributed and accelerated computing facilities. Our map-making framework \texttt{BrahMap} is one such initiative. With \texttt{BrahMap} we strive to provide a user-friendly map-making framework that can be used with a variety of computational facilities, including personal laptops, supercomputing clusters as well as multiple GPUs. In the rest of this paper, we will discuss the implementation details, features, performance scaling, and future scope of \texttt{BrahMap}.


\section{Map-making with \texttt{B\textup{rah}M\textup{ap}}}

The matrices involved in map-making are prohibitively large in size. However, except for the noise covariance for a complex noise model, all the matrices are very sparse in nature. In addition, solving the linear equation \eqref{eq:lineq} with a preconditioned conjugate gradient (PCG) involves only a series of matrix-vector products. As a result, it is enough for us to define the linear operators of map-making not as the arrays of numbers, but as objects with embedded functions to compute the matrix-vector product. Due to the fact that the elements of the matrices can be precomputed in advance, we take advantage of the sparse representation of the linear operators to bring down the number of floating-point operations compared to their dense formalism by several orders of magnitude. In \texttt{BrahMap}, we exploit these properties to offer a scalable and modular \textit{map-making framework}. We provide a number of independent linear operators used in map-making with a user-friendly Python interface. The user may use these operators to perform map-making with GLS or other suitable methods. Due to several limitations of Python routines in handling large data and computation, we perform the heavy matrix-vector multiplication through functions written in C++. The C++ routines exploit vectorization and, MPI, and OpenMP based hybrid parallelization, to offer scalability and performance across multiple CPUs and computing nodes with least latency overhead. In the next subsection, we discuss more about the programming implementation of \texttt{BrahMap}.


\subsection{Programming implementation}

The programming implementation of \texttt{BrahMap} is based on \texttt{COSMOMAP2}\footnote{\url{https://github.com/giuspugl/COSMOMAP2}}. \texttt{BrahMap} provides linear operator classes for the pointing matrix (and its transpose), block-diagonal preconditioner (see equation \eqref{eq:precond}), and several kinds of noise covariance matrix. We define these linear operators with an adopted version of \texttt{LinearOperator} class of SciPy \cite{scipy} sparse linear algebra library (\texttt{scipy.sparse.linalg}\footnote{\url{https://docs.scipy.org/doc/scipy/reference/sparse.linalg.html}}). Linear operators defined with \texttt{LinearOperator} support mathematical operations such as addition, subtraction, multiplication, exponentiation, and inverse implicitly. In addition, they can also be used in the functions defined in \texttt{scipy.sparse.linalg} for matrix norms, solving linear problems, eigenvalue problems, singular value problems, and complete/incomplete LU factorization out-of-the-box. Most importantly, the \texttt{LinearOperator} class allows users to define their own linear operators independently that can integrate with the other operators seamlessly.

Map-making for CMB experiments, in fact, is not just limited to solving GLS. The map-making on real data involves more complex operation such as filtering, gap-treatment, noise estimation, etc. and often involves other iterative algorithms to calibrate and mitigate various kinds of systematics. The facility for users to define new operators independently in modular fashion, and use a wide set of in-built matrix operations makes \texttt{BrahMap} possible to use for map-making in the widest possible range. To the best of our knowledge, among all the publicly available map-making implementations, such a critical feature is unique to \texttt{BrahMap}.

In the following, we discuss the programming implementation of \texttt{BrahMap} in more detail. The code infrastructure of \texttt{BrahMap} can be divided into three layers:

\begin{enumerate}
  \item \textbf{Python user interface:} As mentioned before, the \texttt{BrahMap} user interface is based on an adopted version of \texttt{LinearOperator} class of the SciPy sparse linear algebra library (\texttt{scipy.sparse.linalg}). The operators defined in \texttt{BrahMap} subclass the \texttt{LinearOperator} class with suitable definition and algorithm for matrix-vector (and the transposed-matrix-vector) multiplication. The matrix vector operations are set to evaluate by overloading the multiplication operation for the \texttt{LinearOperator} class. When an instance of such a class is initialized, it simply defines a Python object, it neither allocates the array for the matrices nor performs any computation. Computations are triggered only when the linear operator is multiplied by a vector of compatible length. This approach reduces the memory footprint of the map-making problem significantly.

  Another precaution that helps us reducing the redundant computation is the pre-computation of the elements of the linear operators for the pointing matrix and block-diagonal preconditioner. These elements are essentially a function of the angle $\theta$ and the diagonal of the noise covariance matrix. Since the pointing matrix and the block-diagonal preconditioner are used in every iteration of the PCG, it only makes sense to precompute their elements in advance and pass them to the linear operators as needed.

  Due to prohibitively large data sets used in map-making, in order to work with full data sets, it is necessary to utilize multiple CPUs across multiple computing nodes. We achieve this in \texttt{BrahMap} with MPI and OpenMP based data distribution and parallelization. We make this happen by making every process have its own instance of the linear operator, each handling only a small subset of data available on a given process. These linear operators are, however, defined in a collective sense, so that their action (matrix-vector multiplication) are defined globally. As a result, the distribution of data across multiple processes and parallelized matrix-vector multiplication happens behind the scene. Even though the individual MPI process allocates the Python arrays using the locally available subset of pointing and TODs, they are passed to C++ to perform the actual matrix-vector multiplication as we discuss below.

  \item \textbf{Computationally extensive routines written in C++:} In the very early stage of the development, we had the entire codebase written in Python. The operations were very slow mainly due to lack of vectorization. To overcome this issue, we decided to write the computationally extensive functions in compiled language. For this purpose, we decided to go with C++. Our choice was motivated by two fundamental advantages that C++ could offer: First, exposing the generic Python data types to C++ using C++ templates and second, the ability to control hardware-level optimization. Just with this approach, we were able to gain speedup in performance by the factor of 22.

  Another critical advantage of using C++ backend is that we can also bypass the global interpreter lock (GIL) of Python and introduce the OpenMP based multithreading with higher degree of control in C++ functions. To make parallelization even more robust, \texttt{BrahMap} handles all MPI communications on the C++ side. It is to avoid the overhead associated with pickling and un-pickling of Python objects when MPI communications are performed in Python using libraries such as \texttt{mpi4py}\footnote{\url{https://github.com/mpi4py/mpi4py}}. With the MPI and OpenMP based hybrid parallelization, we were able to scale our code over a large number of CPUs with relatively low memory footprint. We discuss the performance scaling with this approach in the next section.

  \item \textbf{Python bindings for C++ routines:} The Python interface of \texttt{BrahMap} and C++ computing backend are not naturally interoperable. One needs to bind the interface with the C++ function to ensure correct type conversion from one language to another. Among various options available for binding, we opted for \texttt{pybind11} \cite{pybind11}. \texttt{pybind11} is a header-only library that exposes Python data types to C++ and vice versa. It offers native integration with NumPy \cite{numpy} and C++ standard template library (STL) data structures. To prevent any situation of dangling pointers and memory leakage, we opted to allocate (and de-allocate) all the arrays in the Python side and pass them to C++ whenever they were required in computations.

  The core task of the bindings in \texttt{BrahMap} is to expose the NumPy arrays to C++ functions with correct type-conversion with least overhead. \texttt{pybind11} offers a variety of solutions to make this happen. We settled on the solution offered by \texttt{pybind11} that exposes the Python buffer protocol\footnote{\url{https://docs.python.org/3/c-api/buffer.html}} to C++. Then we wrote the function with templates for Python array API implementation, float-point, and integer data types. This way, our binding functions are written in an array API implementation agnostic manner, and it can consume any Python array that complies with Python array API. This includes NumPy, CuPy\footnote{\url{https://github.com/cupy/cupy}} array and a lot more. This implementation performs the type conversion in two steps. In the first step, the \texttt{std::function} takes the Python array object as template argument and passes it to a lambda function as \texttt{pybind11::buffer} object. Then we strip off the array pointer from the buffer and pass it to the C++ function. In a comparative test with other available options, we found that this solution is the most robust in terms of memory and performance overhead.
\end{enumerate}

To summarize, we use an adopted version of \texttt{LinearOperator} class of SciPy sparse linear algebra library to define the linear operators needed in map-making. These linear operator objects allow overloading the multiplication operation to perform matrix-vector products. We delegate the computation of the matrix-vector product to the compiled routines written in C++ where we take full advantage of MPI and OpenMP based hybrid parallelization to offer fast and scalable computation.

\subsection{Performance scaling}

As we mentioned earlier, we started \texttt{BrahMap} as a pure Python library. With a C++ based computational backend compiled with vectorization support, we were able to \textbf{achieve 22 times performance gain}. It brought the performance of our serial GLS implementation to the level of other simple GLS implementations.

\begin{figure}[htbp]
  \subfloat[Strong scaling for GLS implementation in \texttt{BrahMap}\label{fig:strong_scale}]{\centerline{\includegraphics{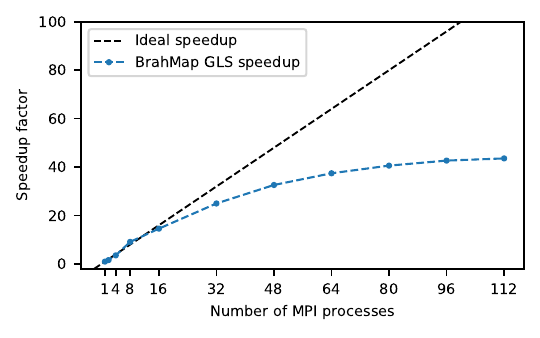}}}\hfill
  \subfloat[Strong scaling for individual operators used in GLS implementation\label{fig:strong_scale_ops}]{\centerline{\includegraphics{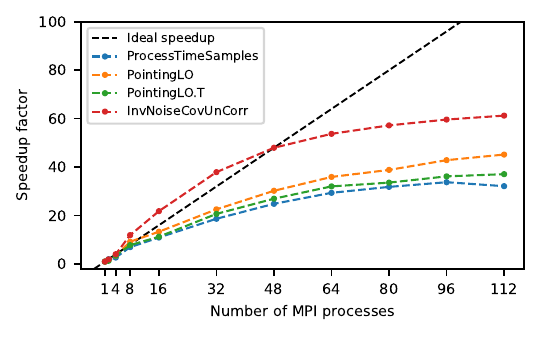}}} 
  \caption{Strong scaling performance. It shows the performance speedup when the number of MPI processes is increased for the problem of fixed size.}
  \label{fig:strong}
\end{figure}

To evaluate the performance of parallelized implementation, we performed the strong and weak scaling test on multiple MPI processes\footnote{We perfomed the scaling tests with Python v3.10, NumPy v1.26, SciPy v1.14, and \texttt{pybind11} v2.11 stack.}. Fig. \ref{fig:strong_scale} shows the strong scaling plot for GLS implemented with \texttt{BrahMap} as the number of MPI processes is increased for map-making with the same amount of data. We performed this test on a compute node with 112 CPUs and 512 GB RAM. We observe that our code scales very well to 32 processes. After that, the speedup goes down gradually, and it saturates for around 80 processes. We observe similar behavior for individual operators we use in GLS implementation, as shown in Fig. \ref{fig:strong_scale_ops}. An exception to this behavior is \texttt{InvNoiseCovUnCorr} which stands for the inverse noise covariance operator. The other operators shown in the figure are \texttt{PointingLO} and \texttt{PointingLO.T} for pointing and transposed-pointing operator respectively. The class \texttt{ProcessTimeSamples} performs the initializations and pre-computations. Once initialized, the instances of all these operators can be re-used many times.


As we increase the total number of MPI processes keeping the global size of the problem constant, the problem size per process gets smaller. The time it takes to perform the parallel computations becomes shorter consequently. It doesn't however reduces the computational cost associated with serial computations and number of MPI communications which starts to dominate over parallel computations for large number of MPI processes. This explains the degradation in strong scaling performance of GLS implementation and individual operators for large number of MPI processes.


\begin{figure}[ht]
  \subfloat[Weak scaling for GLS implementation in \texttt{BrahMap}\label{fig:weak_scale}]{\centerline{\includegraphics{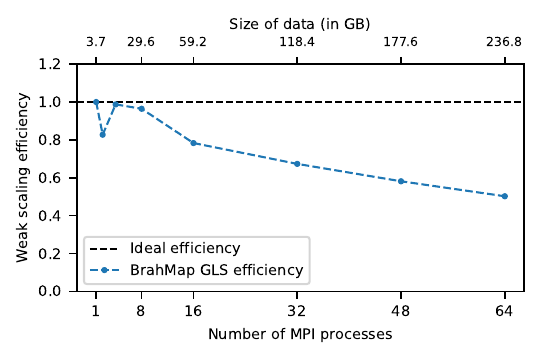}}}\hfill
  \subfloat[Weak scaling for individual operators used in GLS implementation\label{fig:weak_scale_ops}]{\centerline{\includegraphics{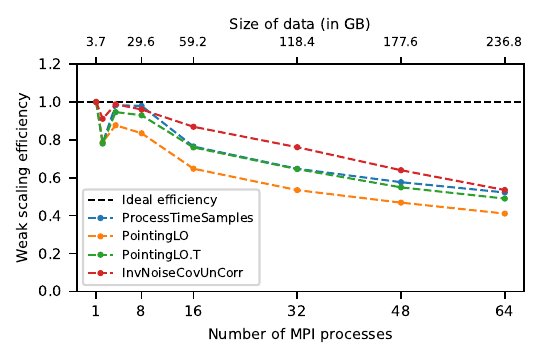}}} 
  \caption{Weak scaling performance. It shows the efficiency of parallelization as the size of problem is increased along with the number of MPI processes used in parallelization.}
  \label{fig:weak}
\end{figure}

Fig. \ref{fig:weak_scale} shows the weak scaling plot for GLS implemented with \texttt{BrahMap}, as the number of MPI processes is increased proportionally with the problem size. We observe that the weak scaling efficiency is very consistent for small number of processes but degrades for large number of MPI processes. The reason behind the loss in efficiency is that the problem becomes more data intensive as the size of the problem is increased. Consequently, the time it takes to perform MPI communication increases, leading to a large MPI communication overhead. This is supported by the weak scaling behavior of individual operators as shown in Fig. \ref{fig:weak_scale_ops}. The \texttt{InvNoiseCovUnCorr} operator, with no MPI communications, offers the best weak scaling, whereas \texttt{PointingLO} operator that performs \texttt{MPI\_Allreduce} operation, exhibits the worst scaling.



In case of \texttt{PointingLO} operator, we perform the reduction operation over an array of size equal to two or three times the number of sky pixels. As we increase the target number of sky pixels, the MPI communication overhead increases as well. The solution to this problem is reducing the number of copies of such an array on a given compute node. To make this happen, as mentioned before, we are using MPI and OpenMP based hybrid parallelization. With this solution, we are targeting to reduce the number of MPI processes per node (while increasing the number of threads per MPI process), without affecting the parallelization performance gain. The OpenMP based parallelization is under the final stage of performance tuning stage, and we are planning to release it soon.


\section{Future extensions to \texttt{B\textup{rah}M\textup{ap}}}

\texttt{BrahMap} is publicly available on Github\footnote{\url{https://github.com/anand-avinash/BrahMap}} with a detailed documentation\footnote{\url{https://anand-avinash.github.io/BrahMap}}. Using the features currently provided by \texttt{BrahMap}, it can be used to perform map-making on TODs for multiple detectors containing white noise. In the TODs coming from real experiment, the noise contribution is, however, complicated, and it also includes instrumental systematics. In order to perform map-making for such a dataset, it becomes important to introduce filtering templates and improved noise covariance models. Filtering template operators are usually sparse; they can be easily defined using \texttt{LinearOperator} class. Noise covariance, on the other hand, is dense. Given the enormous size of the noise covariance, it is impossible to load the entire dense matrix into memory. In such a case, we use some well-structured matrices, very close to the actual noise covariance, to obtain the maximum-likelihood maps using equation \eqref{eq:maxl_solution}. A preferred choice for such structured matrix is the circulant matrix. Map-making with circulant noise covariance has been discussed elsewhere in the literature \cite{mm1, mm2, mm3}. Since a circulant matrix can be diagonalized with the Fourier transform of its first row, we can compute matrix-vector multiplication with its inverse in just $O(n\log n)$ floating-point operations. Our work on circulant noise covariance operator (and its inverse) is in progress, and we plan to include its implementation in our next release.

On the other hand, while MPI and OpenMP based hybrid parallelization in \texttt{BrahMap} is robust, we note that it is possible to improve the performance even more by offloading the computations to graphical processing units (GPUs). Given that most of our compute-intensive functions have a simple loop structure which we parallelize using OpenMP on CPUs, we will be able to use the same pragma directives with additional OpenMP \texttt{target} directives to perform the computations on GPUs. Such an implementation, however, would require memory allocation on GPUs in the Python interface. It can be done using CuPy arrays. As CuPy arrays are based on the Python array API, we can pass them to C++ using the same template binding functions that we discussed before. We plan to release \texttt{BrahMap} with GPU offloading using this approach in the near future.



\section{Conclusion}

The development of an efficient map-making framework is instrumental in tackling the data analysis challenges for future CMB experiments. Our map-making framework, \texttt{BrahMap}, addresses this challenge by offering a general-purpose and high performance solution. Our scalable and modular map-making framework can be used for virtually all map-making methods that involve matrix operations. It provides the user with the ability to define independently the linear operators compatible with the tools that SciPy offers. The user-friendly Python interface allows it to integrate seamlessly with existing simulation tools for the CMB experiments. While offering an intuitive interface, \texttt{BrahMap} tackles the computational challenge by using optimized C++ functions behind the scenes. These functions written with MPI and OpenMP offer parallelization and scalability across multiple CPUs and compute nodes with minimum communication overhead. Moreover, since the binding functions in \texttt{BrahMap} are written in Python array API implementation agnostic manner, it allows the extending \texttt{BrahMap} to utilize CuPy arrays as well with no change in binding functions. With extensive improvements and updates planned in the future, including more realistic noise covariance models and GPU offloading, \texttt{BrahMap} is poised to become an essential tool in the map-making pipeline for the future generation of CMB experiments.

\section*{Acknowledgment}

This work is supported by Italian Research Center on High Performance Computing, Big Data and Quantum Computing (ICSC), project funded by European Union - NextGenerationEU - and National Recovery and Resilience Plan (NRRP) - Mission 4 Component 2 within the activities of Spoke 3 (Astrophysics and Cosmos Observations). This work has also received funding by the European Union’s Horizon 2020 research and innovation program under grant agreement no. 101007633 CMB-Inflate. A.A. is also supported by the InDark and LiteBIRD INFN projects. We thank INFN LiteBIRD project for granting access to the LEONARDO supercomputer, owned by the EuroHPC Joint Undertaking, hosted by CINECA (Italy) and the LEONARDO consortium.




\IEEEtriggeratref{11} 
\printbibliography

\end{document}